\documentclass[12pt]{iopart}

\usepackage{graphicx}
\usepackage{iopams}
\usepackage{cite}
\usepackage{braket}
\begin{document}

\title[Unconventional antiferromagnetic spin fluctuations in Yb$_3$Pt$_4$]{\boldmath Magnetic transition and spin fluctuations in the unconventional antiferromagnetic compound Yb$_3$Pt$_4$}

\author{S. Zhao$^{1}$, D. E. MacLaughlin$^{1}$, O. O. Bernal$^{2}$, J. M. Mackie$^{1}$, C. Marques$^{3,4}$, Y. Janssen$^{3}$, M. C. Aronson$^{3,4}$}

\address{$^{1}$ Department of Physics and Astronomy, University of California, Riverside, CA 92521, U.S.A.}
\address{$^{2}$ Department of Physics and Astronomy, California State University, Los Angeles, CA 90032, U.S.A.}
\address{$^{3}$ Condensed Matter Physics and Materials Science Department, Brookhaven National Laboratory, Upton, New York 11973, U.S.A.}
\address{$^{4}$ Department of Physics and Astronomy, Stony Brook University, Stony Brook, New York 11794-3800, U.S.A.}
\ead{songruiz@ucr.edu}

\begin{abstract}
Muon spin rotation and relaxation measurements have been carried out on the unconventional antiferromagnet Yb$_{3}$Pt$_{4}$. Oscillations are observed below $T_{\mathrm{N}} = 2.22(1)$~K, consistent with the antiferromagnetic (AFM) N\'eel temperature observed in bulk experiments. In agreement with neutron diffraction experiments the oscillation frequency~$\omega_\mu(T)/2\pi$ follows a $S = 1/2$ mean-field temperature dependence, yielding a quasistatic local field 1.71(2)~kOe at $T = 0$. A crude estimate gives an ordered moment of ${\sim}0.66{\mu}_{\mathrm{B}}$ at $T = 0$, comparable to $0.81{\mu}_{\mathrm{B}}$ from neutron diffraction. As $T \rightarrow T_{\mathrm{N}}$ from above the dynamic relaxation rate~$\lambda_{\mathrm{d}}$ exhibits no critical slowing down, consistent with a mean-field transition. In the AFM phase a $T$-linear fit to $\lambda_{\mathrm{d}}(T)$, appropriate to a Fermi liquid, yields highly enhanced values of $\lambda_{\mathrm{d}}/T$ and the Korringa constant~$K_\mu^2T/\lambda_{\mathrm{d}}$, with $K_\mu$ the estimated muon Knight shift. A strong suppression of $\lambda_{\mathrm{d}}$ by applied field is observed in the AFM phase. These properties are consistent with the observed large Sommerfeld-Wilson and Kadowaki-Woods ratios in Yb$_{3}$Pt$_{4}$ (although the data do not discriminate between Fermi-liquid and non-Fermi-liquid states), and suggest strong enhancement of ${\mathbf{q}} \approx 0$ spin correlations between large-Fermi-volume band quasiparticles in the AFM phase of Yb$_{3}$Pt$_{4}$.
\end{abstract}

\submitto{\JPCM}
\maketitle

\section{Introduction}

The term \textit{quantum criticality} refers to the ensemble of phenomena associated with a critical phase transition at zero temperature. For a number of reasons~\cite{Si06} strongly correlated-electron systems in general, and heavy-fermion (HF) metals in particular, are particularly convenient systems in which to study quantum criticality~\cite{Jaramillo:2009fk,Coleman:2005uq,Hussey:2007kx,%
Senthil:2008vn,Si:2001ys,Loehneysen:2007zr,Schofield:2007ly,Gegenwart:2008ve}, since many HF materials possess a quantum critical point (QCP) at $T = 0$. Properties such as unconventional superconductivity, non-Fermi liquid (NFL) behaviour, weak-moment antiferromagnetism, and quasi-ordered phases such as `nematic ordering' in HF systems~\cite{Senthil:2009qf,Knafo:2009bh,Stewart:2001dq,Amato:1997cr} are all intimately related to quantum critical behaviour.

Recent studies of the binary compound Yb$_{3}$Pt$_{4}$~\cite{BSKJ08u,BKSG09,BSKJ09,JKPW10} indicate that it represents a new type of quantum critical system, possessing an unusual magnetic phase diagram and behaviour near the QCP that differs significantly from that of `conventional' HF systems. Yb$_{3}$Pt$_{4}$ crystallises in the rhombohedral Pu$_{3}$Pd$_{4}$-type structure~\cite{CROMER:1973hc,PALENZONA:1977,BENDERSKY:1993ij}, and can be represented by parallel chains of alternating `Yb$_{6}$Pt' octahedra and `Pt$_{6}$Pt' octahedra~\cite{JKPW10,BENDERSKY:1993ij}. Neutron diffraction reveals a $\mathbf{k} = 0$ antiferromagnetic (AFM) structure below a N\'eel temperature~$T_{\mathrm{N}} = 2.4$~K~\cite{JKPW10}, which is also evidenced in transport and magnetic properties measurements~\cite{BSKJ09,BSKJ08u}. The specific heat gives evidence that the AFM transition is mean-field at low applied magnetic fields, evolving into a lambda-like second-order transition at a critical end point (CEP) $T_{\mathrm{CEP}} = $ 1.2~K, $H_{\mathrm{CEP}} = 15.2$~kOe. Finally $T_{\mathrm{N}} \rightarrow 0$ at $H_{\mathrm{QCP}} \approx 16.2$~kOe, indicating a magnetic field tuned AFM QCP\@. The abnormal features related to the QCP in Yb$_{3}$Pt$_{4}$ are: (a) There is no mass divergence at the QCP, and (b) the FL state is not destroyed across the QCP. This indicates the magnetic order might be driven by the exchange enhancement of the Fermi liquid itself in this system and could therefore represent a new route to quantum criticality in heavy-fermion metals.

This paper reports a muon spin rotation and relaxation ($\mu$SR) study of the static and dynamic magnetism of Yb$_{3}$Pt$_{4}$ single crystals. In the time-differential $\mu$SR technique~\cite{ASmuSR,Brew94} spin-polarised positive muons ($\mu^+$) are stopped at the sample and precess in their local fields; the time dependence of the ensemble $\mu^+$ spin polarisation, $P(t)$, is obtained from the asymmetry~$A(t)$ of the decay positron momenta. In Yb$_{3}$Pt$_{4}$ the onset of damped oscillations indicates an AFM transition at $T_{\mathrm{N}} = 2.21(1)$ K, slightly lower than but consistent with the value obtained from bulk data. The oscillation frequency yields a crude estimate for the ordered magnetic moment of ${\sim}0.66\mu_{\mathrm{B}}$/Yb ion, consistent with the value of 0.81~$\mu_{\mathrm{B}}$/Yb ion obtained from neutron diffraction~\cite{JKPW10}. The oscillation frequency follows a $S = 1/2$ mean field temperature dependence, also in agreement with neutron diffraction experiments. 

The dynamic $\mu^{+}$ spin relaxation rate~$\lambda_{\mathrm{d}}(T)$ increases upon approaching $T_{\mathrm{N}}$ from above but shows no sign of a critical divergence, consistent with a mean-field transition. Below $T_{\mathrm{N}}$ $\lambda_{\mathrm{d}}(T)$ drops rapidly from $T_{\mathrm{N}}$ to $\sim$1.8~K, then decreases less rapidly with decreasing temperature, passing through a shallow minimum at $T_{\mathrm{min}} \sim0.1$~K\@. Below $\sim$1.5~K $\lambda_{\mathrm{d}}(T)$ can be fit with a sublinear power law (above $T_{\mathrm{min}}$), suggestive of NFL behaviour, or alternatively a two-component form that assumes Korringa relaxation of $\mu^+$ spins and Yb$^{3+}$ moments by spin scattering of Fermi-liquid quasiparticles. Highly enhanced values are obtained for the coefficient~$\lambda_{\mathrm{d}}/T$ and the Korringa product~$K_\mu^2 T/\lambda_{\mathrm{d}}$, where the $\mu^{+}$ Knight shift~$K_\mu$ is estimated from the magnetic susceptibility. Strong field-induced suppression of $\lambda_{\mathrm{d}}$ is observed below $T_{\mathrm{N}}$ but not in the high-temperature paramagnetic phase. These results are suggestive of strong enhancement of ${\mathbf{q}} \approx 0$ spin correlations in Yb$_{3}$Pt$_{4}$, and confirm the unusual nature of the phase transition in this compound.

\section{Experiments}

$\mu$SR experiments were carried out on Yb$_{3}$Pt$_{4}$ single crystals at the M15 beam line at \mbox{TRIUMF}, Vancouver, Canada. A number of needle-like single crystals (the $c$-axis is the long direction) were aligned on a rectangular silver plate of area $\sim$1~cm$^{2}$ and fastened with GE varnish with the crystalline $c$ axis perpendicular to the beam direction. The experiments were carried out in the top-loading $^{3}$He-$^{4}$He dilution refrigerator at \mbox{TRIUMF} using the standard time-differential $\mu$SR technique~\cite{ASmuSR,Brew94}. Asymmetry data $A(t)$ were taken in longitudinal fields up to 10~kOe at 25~mK, 1.2~K and 2.3~K, and over the temperature range~25~mK--4~K in weak longitudinal fields~15 and 20~Oe. The latter served to `decouple' $^{195}$Pt nuclear dipolar fields~\cite{HUIN79} that otherwise would have made an unwanted contribution to $A(t)$. These fields also affected the Yb$^{3+}$ moment fluctuations, as discussed in \sref{sect:relax}.

Magnetic susceptibility measurements were carried out on a single crystal from the $\mu$SR sample in a Quantum Design MPMS magnetometer for applied fields of 100 Oe, 10~kOe, and 20~kOe over the temperature range~2--300~K\@. The crystal was aligned with the field perpendicular to the $c$ axis. The data (not shown) agree within errors with previously-published results~\cite{BKSG09,BSKJ09}.

\section{Results and discussion} 

\begin{figure}[bht]
\begin{center} 
\includegraphics[clip=,width=0.55\textwidth]{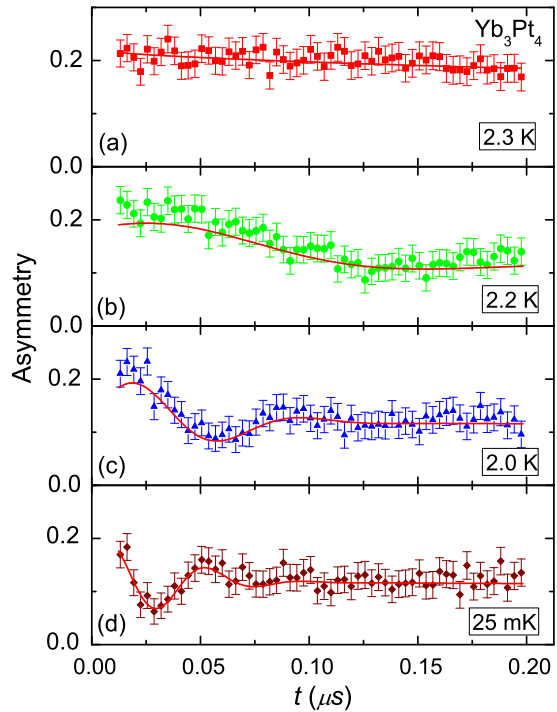}
\end{center} 
\caption{\label{Figure1} Representative early-time $\mu$SR asymmetry data in Yb$_{3}$Pt$_{4}$ at selected temperatures. Solid curves: fits as described in text.}
\label{fig:earlytimeasy}
\end{figure}

Representative early-time asymmetry plots $A(t)$ are shown in \fref{fig:earlytimeasy}. 
At low temperatures the data exhibit damped oscillation due to precession of $\mu^{+}$ spins in the spontaneous local field~${\mathbf{B}}_{\mathrm{loc}}$ due to the ordered Yb$^{3+}$ moments. The late-time $A(t)$ data (not shown) are from $\mu^{+}$ spin components parallel to ${\mathbf{B}}_{\mathrm{loc}}$~\cite{HUIN79}, and exhibit dynamic (spin-lattice) relaxation due to thermal fluctuations of the $\mu^{+}$ local field.

To model this situation the following two-component function was fit to the asymmetry data for $T < T_{\mathrm{N}}$:
\begin{equation}
A(t) = A_{\mathrm{s}}\exp(-\Lambda_{\mathrm{s}}t) \cos(\omega_{\mu}t) + A_{\mathrm{d}}\exp[-(\lambda_{\mathrm{d}}t)^{K}] + A_{\mathrm{bkg}}\,.
\label{eq:fitfunction}
\end{equation}
The first term describes the early-time oscillation in a static field with an exponential damping envelope, and the second term is the `stretched exponential' often used to describe an inhomogeneous distribution of dynamic $\mu^{+}$ relaxation rates~\cite{Yaouanc:2005fu}. Here $\omega_{\mu}$ is the angular frequency of the $\mu^{+}$ spin precession, $\Lambda_{\mathrm{s}}$ is the exponential damping rate, $\lambda_{\mathrm{d}}$ is the dynamic relaxation rate, and $K < 1$ is the `stretching power'. The third term represents a constant background signal arising from muons that stop in the silver sample holder. The total initial asymmetry~$A(0) = A_{\mathrm{s}} + A_{\mathrm{d}} + A_{\mathrm{bkg}}$ is assumed to be independent of temperature and applied field. The fitting results based on \eref{eq:fitfunction} are shown in \fref{fig:earlytimeasy} as solid curves. \Eref{eq:fitfunction} is also expected to model $A(t)$ data for $T > T_{\mathrm{N}}$ with $A_{\mathrm{s}} = 0$.

\subsection{$\mu^{+}$ spin precession}

\begin{figure}[bht]
\begin{center}
\includegraphics[clip=,width=0.55\textwidth]{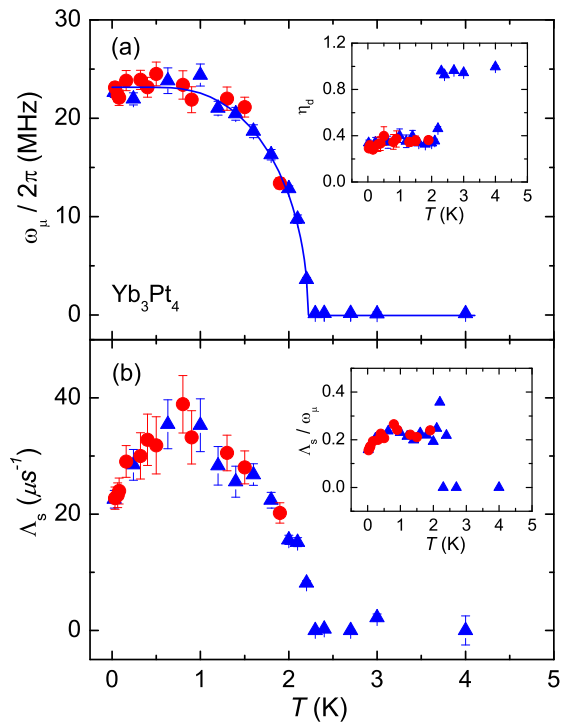}
\end{center}
\caption{Temperature dependencies of (a)~$\mu^{+}$ spin precession frequency~$\omega_\mu$ and (b)~exponential damping rate~$\Lambda_{\mathrm{s}}$ in Yb$_{3}$Pt$_{4}$. Triangles: longitudinal field~$H = 15$~Oe. Circles: $H = 20$~Oe. Curve: $S = 1/2$ mean-field model fit to $\omega_\mu(T)$. Inset to (a): late-time fraction~$\eta_{\mathrm{d}}$. Inset to (b): ratio~$\Lambda_{\mathrm{s}}/\omega_\mu$.}
\label{fig:earlytimeparams}
\end{figure}

The temperature dependence of $\omega_{\mu}$ is shown in \fref{fig:earlytimeparams}(a). 
As for the neutron-diffraction results~\cite{JKPW10}, $\omega_\mu(T)$ is well described by a $S = 1/2$ mean-field model [curve in \fref{fig:earlytimeparams}(a)]. A fit to this function yields $T_{\mathrm{N}} = 2.22(1)$~K, which is slightly lower than the value 
 obtained from susceptibility and specific heat measurements~\cite{BKSG09}. The zero-temperature frequency~$\omega_\mu(0)/2\pi = 23.2(3)$~MHz corresponds to a quasistatic local field $B_{\mathrm{loc}}(0) = \omega_{\mu}(0)/\gamma_{\mu} = 1.71(2)$~kOe. 

A very rough estimate of the static Yb$^{3+}$ moment can be obtained by taking the magnetisation~$M$ due to uniformly-polarised Yb$^{3+}$ moments to be an estimate~$B_{\mathrm{loc}}^{\mathrm{est}}$ of the local dipolar field at the $\mu^{+}$ stopping site: $B_{\mathrm{loc}}^{\mathrm{est}} = 4\pi M$\@. For Yb$_{3}$Pt$_{4}$ the unit cell volume $v_c = 811.65$~\AA$^3$, and a unit cell contains 18 Yb ions~\cite{BKSG09}. Thus $M = 18\mu_{\mathrm{B}}/v_c = 206~\mathrm{emu}/\mu_{\mathrm{B}}$, so that $B_{\mathrm{loc}}^{\mathrm{est}} = 2.58~\mathrm{kG}/\mu_{\mathrm{B}}$. Comparison of the measured field with this value yields an estimated moment of ${\sim}0.66\mu_{\mathrm{B}}$/Yb ion, in better-than-expected agreement with the neutron diffraction result~$0.81(5)\mu_{\mathrm{B}}$/Yb ion~\cite{JKPW10}. Obviously this procedure takes no account of any transferred hyperfine contribution to $B_{\mathrm{loc}}$, the specific $\mu^{+}$ site, or the Yb$^{3+}$ moment configuration in the AFM phase; it most likely overestimates $B_{\mathrm{loc}}^{\mathrm{est}}$. A more accurate estimate would use a lattice-sum calculation of the dipolar field at the $\mu^{+}$ stopping site from the Yb$^{3+}$ magnetic structure determined by neutron diffraction; unfortunately this cannot be done at present because the stopping site is not known.

The late-time fraction~$\eta_{\mathrm{d}}$, defined by
\begin{eqnarray}
\eta_{\mathrm{d}} = \frac{A_{\mathrm{d}}}{A_{\mathrm{s}}+A_{\mathrm{d}}} \,,
\end{eqnarray} 
is given in the inset of \fref{fig:earlytimeparams}(a). The step-like change indicates the disappearance of the quasistatic component of ${\mathbf{B}}_{\mathrm {loc}}$ above $T_{\mathrm{N}}$, consistent with the evidence for a sharp mean-field-like transition obtained from $\omega_{\mu}(T)$. The sharpness of the step indicates that the spread in $T_{\mathrm{N}}$ is small.

The exponential damping rate~$\Lambda_{\mathrm{s}}$ is shown in \fref{fig:earlytimeparams}(b). We shall see below that the dynamic relaxation rate is much smaller than $\Lambda_{\mathrm{s}}$; the latter is therefore dominated by a static inhomogeneous distribution of $\mu^{+}$ precession frequencies~\cite{ASmuSR}. The decrease of $\Lambda_{\mathrm{s}}$ at low temperatures is not understood. The inset to \fref{fig:earlytimeparams}(b) gives the ratio~$\Lambda_{\mathrm{s}}/\omega_\mu$. Except for a spike at $T_{\mathrm{N}}$, most likely due to a narrow spread of transition temperatures, this ratio is fairly small (20--30\%); the spontaneous field in the AFM phase is relatively homogeneous.

\subsection{$\mu^{+}$ spin relaxation}\label{sect:relax}

\begin{figure}[bht]
\begin{center}
\includegraphics[clip=,width=0.55\textwidth]{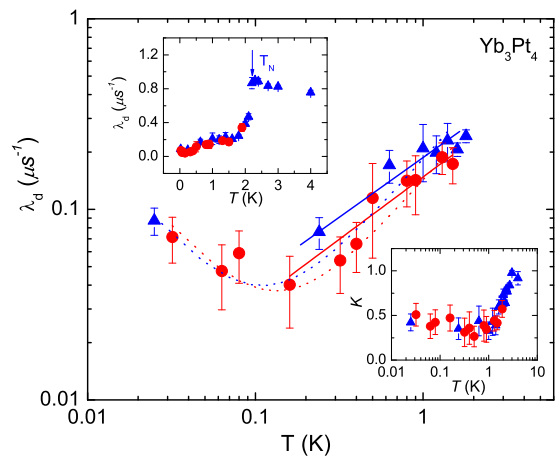}
\end{center}
\caption{(Colour online) Temperature dependence of dynamic $\mu^{+}$ spin relaxation rate~$\lambda_{\mathrm{d}}$ in Yb$_{3}$Pt$_{4}$, $0.025~\mathrm{K} \le T \le 1.8$~K\@. Triangles: longitudinal field~$H = 15$~Oe. Circles: $H = 20$~Oe. Solid lines: fits to the power law~$\lambda_{\mathrm{d}} \propto T^\alpha$. Dashed curves: fits to $\lambda_{\mathrm{d}} = AT + B/T$. Upper left inset: $\lambda_{\mathrm{d}}(T)$, $0.025~\mathrm{K} \le T \le 4$~K\@. Lower right inset: stretching power~$K$.}
\label{fig:dynamicrelax}
\end{figure}

The temperature dependence of the dynamic $\mu^{+}$ spin relaxation rate~$\lambda_{\mathrm{d}}$ in longitudinal fields~$H$ of 15~Oe and 20~Oe is plotted in \fref{fig:dynamicrelax}. 
Even for such small fields there is a visible field dependence, discussed in more detail below. Below $T_{\mathrm{N}}$ $\lambda_{\mathrm{d}}$ drops rapidly with decreasing temperature, suggesting the opening of a spin gap. For $T \lesssim 1.8$~K $\lambda_{\mathrm{d}}$ decreases considerably less rapidly, passing through a shallow minimum at $\sim$0.1~K and then increasing, by a factor of 2 at 0.025~K\@. 

Fits to (i) a power law
\begin{equation}
\lambda_{\mathrm{d}} \propto T^\alpha
\label{eq:powerlaw}
\end{equation}
between 0.2~K and 1.8~K and (ii)~the two-component form
\begin{equation}
\lambda_{\mathrm{d}} = AT + B/T \,,
\label{eq:Korringa}
\end{equation}
down to 25~mK are both plotted in \fref{fig:dynamicrelax}. Parameters from the fits are given in \tref{tab:params}.
\begin{table}[ht]
\caption{\label{tab:params}Parameters from fits of \eref{eq:powerlaw} and \eref{eq:Korringa} to Yb$_3$Pt$_4$ $\mu^+$ spin relaxation data (\protect\fref{fig:dynamicrelax}). See text for details.}
\begin{indented}
\item[]\begin{tabular}{@{}lll}
\br
$H$ & 15 Oe & 20 Oe \\
\mr
$\alpha$ & 0.54(15) & 0.76(9) \\
$A$ ($\mu\mathrm{s} ^{-1}\ \mathrm{K}^{-1}$) & 0.18(2) & 0.16(1) \\
$B$ ($\mu\mathrm{s} ^{-1}\ \mathrm{K}$) & 0.0022(9) & 0.0024(4) \\
\br
\end{tabular}
\end{indented}
\end{table}
The power-law and two-component fits are of comparable quality, and it is not possible to distinguish between them from the data. 

A sublinear power law and/or low-temperature increase might suggest NFL behaviour, although the former does not explain the latter. \Eref{eq:Korringa} can, however, be interpreted as a model for $\mu^{+}$ relaxation by two mechanisms, both of which involve quasiparticle excitations in a Fermi liquid. These excitations couple directly to $\mu^+$ spins [the Korringa mechanism~\cite{Korr50}, first term in \eref{eq:Korringa}], and to localised Yb$^{3+}$ moment fluctuations, also via a Korringa mechanism, which in turn contribute to $\mu^+$ spin relaxation [second term in \eref{eq:Korringa}]. For a Fermi liquid the relaxation is linear in $T$ in both cases. The Korringa relaxation rate of Yb$^{3+}$ moments decreases with decreasing temperature, leading to transfer of fluctuation noise power to low frequencies. This results in an inverse-$T$ contribution to $\lambda_{\mathrm{d}}$, which samples the noise power at the $\mu^{+}$ Zeeman frequency. 

There is evidence for a Fermi liquid in the AFM phase of Yb$_{3}$Pt$_{4}$~\cite{BSKJ09}, so that we consider further consequences of the Fermi-liquid \textit{ansatz}. The Korringa product~\cite{Korr50}
\begin{equation}
{\cal S}_\mu = K_\mu^2\,T/\lambda_{\mathrm{d}}
\end{equation}
relates the $\mu^{+}$ Knight shift~$K_\mu$ and the dynamic relaxation rate in a Fermi liquid. For a non-interacting Fermi gas
\begin{equation}
{\cal S}_\mu = {\cal S}_\mu^0 =\frac{\hbar}{4\pi k_{\mathrm{B}}} \left(\frac{\gamma_e}{\gamma_\mu}\right)^2 = 2.593 \times 10^{-8}~\mathrm{s~K} \,,
\end{equation}
where $\gamma_e$ and $\gamma_\mu$ are the electron and muon gyromagnetic ratios, respectively~\cite{Korr50}. The coupling strength, density of band states, etc., cancel in the product. The ratio ${\cal S}_\mu/{\cal S}_\mu^0$ is affected, sometimes strongly, by details of the coupling and electron correlations~\cite{Mori63}, but is expected to remain within an order of magnitude or so of unity. 

Unfortunately $K_\mu$ cannot be easily measured in the AFM phase. In its absence we use the relation~$K_\mu = a_\mu^{\mathrm{est}} \chi_{\mathrm{mol}}$~\cite{Slic96}, where $a_\mu^{\mathrm{est}} = B_{\mathrm{loc}}^{\mathrm{est}}/N\mu_{\mathrm{B}}$ is the estimated $\mu^+$-spin/Yb$^{3+}$-moment coupling constant and $\chi_{\mathrm{mol}}$ is the molar susceptibility. Using the experimental values $\chi_{\mathrm{mol}}(T_{\mathrm{N}}) = 0.58$~emu/mol Yb~\cite{BKSG09} and $T/\lambda_{\mathrm{d}} = 1/A = 5.6 \times 10^{-6}$~s~K ($H = 15$ Oe) from \tref{tab:params}, with $a_\mu^{\mathrm{est}} = 0.46$~mol~emu$^{-1}$ we find 
\begin{equation}
{\cal S}_\mu \approx 5.0 \times 10^{-7}~\mathrm{s~K} \approx 15\, {\cal S}_\mu^0\,.
\end{equation}
The rough agreement suggests that itinerant electrons in Yb$_3$Pt$_4$ are indeed in a Fermi-liquid state in the AFM phase at intermediate temperatures. Better agreement cannot be expected for a number of reasons, including the stretched-exponential (inhomogeneous) nature of the relaxation function, errors in the estimation of $B_{\mathrm{loc}}$, effective $\mathrm{spin} \ne 1/2$, and the correlated nature of the conduction-electron band. 

We particularly note the effect of enhancement of the complex susceptibility~$\chi(\mathbf{q},\omega)$ near $\mathbf{q} = 0$, which increases ${\cal S}_\mu/{\cal S}_\mu^0$~\cite{Mori63}. The large Sommerfeld-Wilson and Kadowaki-Woods ratios in the AFM phase~\cite{BSKJ09} are also evidence for such enhancement, and the large Kadowaki-Woods ratio in particular associates this with the band quasiparticles involved in resistive scattering.

The low specific-heat coefficient~$\gamma \approx 0.05$~J mol$^{-1}$ K$^{-2}$ in Yb$_{3}$Pt$_{4}$ at low fields~\cite{BSKJ09} suggests that the Fermi-liquid band electrons are not particularly heavy. For a Fermi liquid a crude upper bound on $\lambda_{\mathrm{d}}$ is given by the general expression
\begin{equation}
\lambda_{\mathrm{d}} \lesssim T (\gamma_\mu B_{\mathrm{loc}})^2 \frac{\hbar}{k_{\mathrm{B}}T^{\ast\,2}} \,,
\end{equation}
where $T^{\ast}$ is an effective `Fermi' temperature~\cite{Slic96}. This yields a very low value of $T^{\ast} \lesssim 1.3$~K, suggesting strong enhancement of spin fluctuations sampled by $\lambda_{\mathrm{d}}$. We conclude that in Yb$_{3}$Pt$_{4}$ both the uniform response and local magnetic fluctuations are anomalously strong at low frequencies in a way that is not reflected in the specific heat; this is of course already indicated by the anomalously large Wilson ratio.

The upper left insert of \fref{fig:dynamicrelax} gives $\lambda_{\mathrm{d}}(T)$ between 25~mK and 4~K\@. It can be seen that in the paramagnetic phase between $T_{\mathrm{N}}$ and 4~K $\lambda_{\mathrm{d}}$ increases with decreasing temperature, with a cusp at $T_{\mathrm{N}}$. The increase is slow, and shows no sign of the divergence at $T_{\mathrm{N}}$ often observed in other systems~\cite{MacLaughlin:2008pi,Yaouanc:2008lh,%
Yaouanc:2005fu,Carretta:2009qo,Dunsiger:2006tw}. Such a divergence is due to critical slowing down of spin fluctuations, which moves fluctuation noise power down to the low frequencies sampled by $\mu$SR\@. Thus the absence of a divergence is consistent with a continuous mean-field-like transition with a very narrow critical width. 

The lower right inset of \fref{fig:dynamicrelax} shows the stretching power~$K$ associated with $\lambda_{\mathrm{d}}$. Below $T_{\mathrm{N}}$ $K \approx 0.4$, indicating considerable inhomogeneity in the relaxation rate. There is only a weak temperature dependence below $\sim$1~K, suggesting the stability of the fitting function in the AFM phase. Above $T_{\mathrm{N}}$ $K \rightarrow 1$, i.e., the relaxation is exponential as expected in a homogeneous paramagnetic state. 

\begin{figure}[hbt]
\begin{center}
\includegraphics[clip=,width=0.55\textwidth]{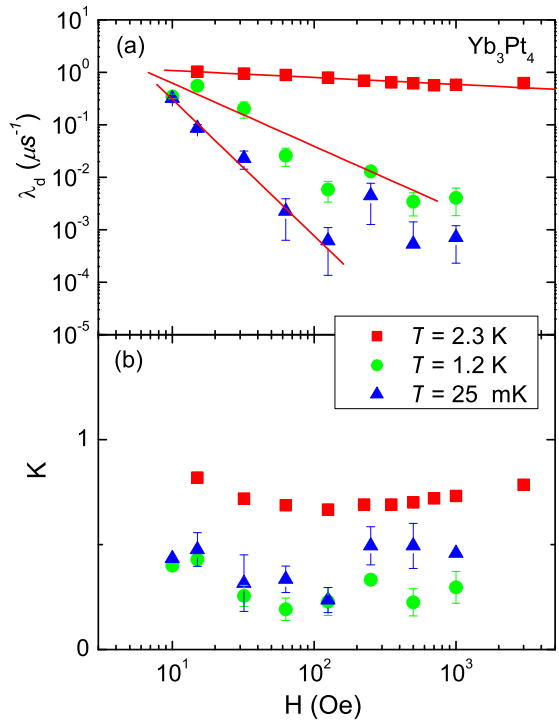}
\end{center}
\caption{(Colour online) Dependence of (a)~dynamic $\mu^{+}$ spin relaxation rate~$\lambda_{\mathrm{d}}$ and (b)~Stretching power~$K$ on applied longitudinal field~$H$ in Yb$_3$Pt$_4$ from fits to stretched-exponential form~$\exp[-(\lambda_{\mathrm{d}}t)^K$.}
\label{fig:fielddep}
\end{figure}

\Fref{fig:fielddep}(a) shows the field dependence of $\lambda_{\mathrm{d}}$ at 25~mK, 1.2~K, and 2.3~K\@. 
It can be seen that at 25~mK $\lambda_{\mathrm{d}}$ is strongly suppressed by the applied field, dropping by an order of magnitude in fields of a few tens of Oe. Above 1~kOe $\lambda_{\mathrm{d}}$ is negligible. The measured rates in the range 0.1--1~kOe are not reliable, being so slow that small systematic errors are important. Fits to the power law~$\lambda_{\mathrm{d}} \propto H^{-\gamma}$ yield $\gamma = 0.13$, 1.2, and 2.6 for 2.3~K, 1.2~K, and 25~mK, respectively. The suppression becomes quite small above $T_{\mathrm{N}}$. Field-induced rate suppression is often a sign of `decoupling' of the muon spin from a distribution of static internal fields~\cite{HUIN79}. This cannot be the case in the present situation, however: in the AFM phase the internal field is strong ($\sim$1.6~kOe), and would not be decoupled by fields less than ten times this value. 

A strong power-law field dependence of the dynamic relaxation is often observed in spin glasses, paramagnetic NFL materials, and geometrically frustrated magnets. It can be evidence for a divergence in the fluctuation noise power spectrum~$J(\omega)$ as $\omega \rightarrow 0$; the relaxation rate, proportional to $J(\omega{=}\omega_\mu)$, tracks the noise spectrum as the field and hence the $\mu^{+}$ Zeeman frequency~$\omega_\mu$ is varied~\cite{MHBI04}. In Yb$_3$Pt$_4$ the origin of the field dependence is not such a divergence: in a magnetically ordered phase the spontaneous local field~$B_{\mathrm{loc}}$ dominates, and $\omega_\mu$ is essentially unchanged until the applied field~$H \approx B_{\mathrm{loc}}$. The observed field dependence is not well understood, but reflects suppression of low-frequency fluctuations by the uniform field. Such suppression would be particularly effective for fluctuations with $\mathbf{q} \approx 0$, consistent with the previously-discussed evidence for correlations with small {\bf q}.

Below $T_{\mathrm{N}}$ the stretching power~$K$ is 0.3-0.5 for all fields, i.e.\ considerably smaller than 1, indicating strong inhomogeneity of the relaxation process. Above $T_{\mathrm{N}}$ $K < 1$ is observed in an applied field, extrapolating to 1 as $H \rightarrow 0$. Here the applied field seems to induce an inhomogeneity similar to (but not as strong as) that in the AFM phase.

\section{Conclusions}

We have carried out a $\mu$SR study of single crystals of the unconventional antiferromagnet~Yb$_{3}$Pt$_{4}$. Well-defined $\mu^{+}$ spin precession is observed in weak-field $\mu$SR data below a N\'eel $\mathrm{temperature} \approx 2.2$~K, in reasonable agreement with previous results for $T_{\mathrm{N}}$~\cite{BKSG09,BSKJ09}. The corresponding quasistatic field yields yields a rough estimate of the ordered Yb$^{3+}$ moment of $0.66\mu_{\mathrm{B}}$, in agreement with the value~$0.81\mu_{\mathrm{B}}$ from neutron diffraction experiments~\cite{JKPW10}. A $S = 1/2$ mean-field temperature dependence of the ordered magnetisation is obtained from both $\mu$SR and neutron diffraction.

In low applied field the temperature dependence of the dynamic $\mu^{+}$ spin relaxation rate~$\lambda_{\mathrm{d}}$ is consistent with $T$-linear Fermi-liquid behaviour over much of the AFM phase. Enhanced magnitudes of both $\lambda_{\mathrm{d}}/T$ and the Korringa constant are observed, and $\lambda_{\mathrm{d}}$ is rapidly suppressed by applied field in the AFM phase. These properties strongly suggest enhancement of spin correlations near $\mathbf{q} = 0$. Below ${\sim}0.05T_{\mathrm{N}}$ additional $\mu^{+}$ spin relaxation is observed, possibly due to \mbox{Yb$^{3+}$-moment} fluctuations due to Korringa scattering. Above $T_{\mathrm{N}}$ $\lambda_{\mathrm{d}}$ exhibits a weak increase with decreasing temperature but no sign of a critical divergence, consistent with a continuous mean-field transition. 

The $\mu$SR evidence for some form of Fermi-liquid behaviour (large Fermi volume) in the AFM phase of Yb$_{3}$Pt$_{4}$, with enhanced spin correlations in the neighbourhood of ${\mathbf{q}} = 0$, seems in conflict with a previous conclusion~\cite{JKPW10}, based on the small specific-heat coefficient, the relatively large ordered Yb$^{3+}$ moment, and well-defined crystalline-field (CEF) excitations, that the Yb$^{3+}$ electrons are localised (small Fermi volume). It should be noted that observation of CEF excitations, with or without magnetic ordering, is fairly common in heavy-fermion metals~\cite{GrSt91}, as are AFM phases with large ordered moments~\cite{Robi00}. We know of no other case of coexistence of large ordered moments with a strongly spin-enhanced Fermi liquid, however. 

In Yb$_3$Pt$_4$ the enhanced magnitude of $\lambda_{\mathrm{d}}$ is not associated with heavy quasiparticles, since the specific heat coefficient shows no such enhancement. This difference, like the large Sommerfeld-Wilson and Kadowaki-Woods ratios, indicates that most of the entropy is removed at or just below the AFM transition but that significant low-frequency spin-fluctuation degrees of freedom remain at low temperatures to contribute to these ratios and $\lambda_{\mathrm{d}}$. Further work will be required to understand this unusual compound in detail.

\section*{Acknowledgement}
We are grateful for the generous technical assistance from the CMMS staff at \mbox{TRIUMF}\@. We would like to thank D A Rose, W Beyermann, and J Morales for their help with the measurements. This work was supported by the U.S. NSF, Grants 0801407 (Riverside), 0604105 (Los Angeles), and 0405961 (Stony Brook). 

\section*{References}




\providecommand{\noopsort}[1]{}\providecommand{\singleletter}[1]{#1}%
\providecommand{\newblock}{}

\end{document}